\documentclass[submission,copyright,creativecommons]{eptcs}
\providecommand{\event}{TARK 2025}
\usepackage{balance}
\usepackage{footmisc}
\usepackage{amsmath}
\usepackage{graphicx}
\usepackage{booktabs}
\usepackage{rotating}
\usepackage{amsthm}
\usepackage{amsfonts}

\usepackage{nicefrac}
\usepackage{xcolor}

\newtheorem{definition}{Definition}
\newtheorem{remark}[definition]{Remark}
\newtheorem{example}[definition]{Example}
\usepackage{subcaption}

\usepackage{iftex}
\ifpdf
  \usepackage{underscore}         % Only needed if you use pdflatex.
  \usepackage[T1]{fontenc}        % Recommended with pdflatex
\else
  \usepackage{breakurl}           % Not needed if you use pdflatex only.
\fi

\title{AI-Generated Compromises for Coalition Formation: Modeling, Simulation, and a Textual Case Study}
\author{
Eyal Briman
\institute{Ben-Gurion University of the Negev\\ Beer Sheva, Israel}
\email{briman@post.bgu.ac.il}
\and
Ehud Shapiro
\institute{Weizmann Institute of Science\\ Rehovot, Israel}
\email{udi.shapiro@gmail.com}
\and
Nimrod Talmon
\institute{Ben-Gurion University of the Negev\\ Beer Sheva, Israel}
\email{nimrodtalmon77@gmail.com}
}
\def\titlerunning{AI-Generated Compromises for Coalition Formation}
\def\authorrunning{E. Briman, E. Shapiro \& N. Talmon}
\begin{document}
\maketitle

\begin{abstract}

The challenge of finding compromises between agent proposals is fundamental to AI sub-fields such as \emph{argumentation}~\cite{rosenfeld2016strategical}, \emph{mediation}~\cite{maturana1996multi}, and \emph{negotiation}~\cite{kraus1997negotiation}. Building on this tradition, Elkind et al.~\cite{elkind2021united} introduced a process for coalition formation that seeks majority-supported proposals preferable to the status quo, using a metric space where each agent has an ideal point. The crucial step in this iterative process involves identifying \emph{compromise proposals} around which agent coalitions can unite. How to effectively find such compromise proposals, however, remains an open question.
We address this gap by formalizing a holistic model that encompasses agent bounded rationality and uncertainty and developing AI models to generate such compromise proposals.

We focus on the domain of collaboratively writing text documents -- e.g., to enable the democratic creation of a community constitution. We apply NLP (Natural Language Processing~\cite{nlp}) techniques and utilize LLMs (Large Language Models~\cite{llm}) to create a semantic metric space for text and develop algorithms to suggest suitable compromise points. To evaluate the effectiveness of our algorithms, we simulate various coalition formation processes and demonstrate the potential of AI to facilitate large-scale democratic text editing, such as collaboratively drafting a constitution—an area where traditional tools are limited.
\end{abstract}

\section{Introduction}\label{sec:intro}

We propose a framework for iterative \emph{compromise-based coalition formation} that enables a set of agents to collaboratively develop a single text document. Each agent starts with an \emph{ideal document}, modeled as a point in a (potentially high-dimensional) metric space. At each step, certain agents may collectively switch to a newly proposed \emph{compromise document}—also represented as a point in the metric space.

Our work generalizes the model of Elkind et al.~\cite{elkind2021united}, who examine an iterative coalition formation process wherein agents only move to a new coalition (i) if its compromise document is closer to their ideal points than the status quo, and (ii) if the new coalition is at least as large as their current one. These two conditions ensure certain theoretical guarantees (e.g., on the convergence of the process; see footnote~\ref{footone} for a formal description of the generalization). However, in many realistic scenarios, agents may not behave strictly according to these criteria. E.g., an agent might rationally move to a smaller coalition if it yields a document that more strongly aligns with its preferences, or it might stochastically deviate from strict rationality due to partial information, uncertainty, or other behavioral considerations.

\paragraph{Generalizing Agent Behavior.}
In contrast to Elkind et al.~\cite{elkind2021united}, we allow for more flexible coalition formation. Agents may:
\begin{itemize}
    \item Move to a new coalition even if that coalition is smaller than their current one,
    \item Take actions probabilistically, representing bounded rationality or incomplete knowledge,

\end{itemize}
These relaxed conditions capture a broader range of real-world behaviors. Thus, our modeling goal is to develop a framework robust enough to accommodate both purely rational and partially rational agents, while still facilitating majority-supported text generation.

\paragraph{Collaborative Text Editing and the Mediator Concept.}
Although Elkind et al.~\cite{elkind2021united} discuss the theoretical dynamics of forming coalitions via compromise proposals, they do not specify \emph{how} such proposals are generated. We address this gap by introducing the notion of an \emph{mediator} that systematically produces compromise documents. Specifically, we embed potential texts in a semantic metric space and employ modern natural language processing (NLP) techniques to measure distances between documents, thereby identifying compromise points that better align with multiple agents’ preferences. This approach is particularly relevant for large-scale, democratic text editing tasks—such as drafting a constitution for a decentralized autonomous organization (DAO)~\cite{dao}—which existing collaborative platforms (e.g., Google Docs, Notion, Wikipedia) do not address in a structured, consensus-driven manner. By modeling documents as points in a high-dimensional embedding space, the mediator can propose new drafts that balance diverse viewpoints, thus paving the way for a more democratic editing process. 
 
\paragraph{Main Contributions.}
Our contributions can be summarized as follows:
\begin{enumerate}
    \item \textbf{Generalizing the Coalition-Formation Model.} We extend the work of Elkind et al.~\cite{elkind2021united} by permitting less restrictive movement rules, thus supporting bounded rationality and agents who may move to smaller coalitions.
    \item \textbf{AI-Mediated Proposal Generation.} We introduce algorithms that employ large language models (LLMs) and other NLP tools to embed and manipulate text documents in a semantic metric space, enabling the discovery of meaningful compromise drafts.
    \item \textbf{Empirical Evaluation in Euclidean and Textual Domains.} We present simulations in both a simplified 2D Euclidean space and a more realistic text-editing environment. Our findings show that—even under relaxed decision rules—agents converge to a majority-supported document that improves upon the status quo.
\end{enumerate}
For space considerations, some text is deferred to the appendix: a more detailed exposition of certain related work; application of the model to the Euclidean setting; more examples; and some details regarding the simulation results.

\subsection{Model State}\label{section:formal static}
The model is defined by the following fixed components\footnote{This is a centralized description for ease of presentation. Conceptually, we envision a decentralized setting, where the mediator operates as a non-centralized tool available to individual coalitions. That is, coalitions may grow in a bottom-up manner, each using an instance of the mediator independently.}:

\begin{itemize}
    \item A metric space \( X \).

    \item A distance function \( \mathbf{d} : X \times X \to \mathbb{R}_{\ge 0} \) defining the metric on \( X \).

    \item A point \( r \in X \), representing the fixed status quo.
    
    \item A set \( V = \{v_1, \ldots, v_n\} \) of \( n \) agents. Each agent \( v \in V \) is associated with an ideal point \( p^v \in X \) and has \emph{Euclidean preferences}~\cite{bogomolnaia2007euclidean}, meaning that preferences are determined by distance from the ideal point.
\end{itemize}

The \emph{state} of the process is given by a coalition structure:
\begin{itemize}
    \item A set \( D = \{d_1, \ldots, d_z\} \), where each \( d_i = (C_i, p_i) \in D \) for \( i \in [z] := \{1, \ldots, z\} \), is a \emph{coalition}. Here, \( C_i \subseteq V \) denotes the set of agents in the coalition and \( p_i \in X \) the compromise point around which the coalition is formed. The coalition structure \( D \) is a partition of the agents: for all \( i \ne j \in [z] \), we have \( C_i \cap C_j = \emptyset \), and \( \bigcup_{i \in [z]} C_i = V \).

We assume \( z \in [n] := \{1, \ldots, n\} \); that is, the number of coalitions does not exceed the number of agents. The notation \( [z] \) refers to the index set of the current coalition structure.
\end{itemize}

\subsection{Initialization, Iterative Process, and Halting Conditions}\label{section:formal process}

Next, we describe a specific modeling and configuration. This approach allows us to present the capabilities of the model in a clear, specific, and traceable manner, making it easier to understand. By focusing on a concrete example, we aim to illustrate the potential applications and advantages of the model, while leaving room to discuss its broader possibilities in the outlook section.

\paragraph{Initialization}

Initially, the process starts with each agent forming its own singleton coalition: namely, $D = \{d_1, \ldots, d_n\} = \{(C_1, p_1), \ldots, (C_n, p_n)\}$ with $C_i = \{v_i\}$ and $p_i \in X$.

\paragraph{Process}

The model contains an entity---the \textbf{mediator}---which is the workhorse of the process.

\begin{definition}[mediator]
A \emph{mediator} $M$ is a function that gets as input a coalition structure $D$ and returns as output a tuple $(d_i, d_j, p)$ with $d_i, d_j \in D$ and $p$ a point in the metric space.
\end{definition}

Intuitively, the mediator suggests that two coalitions, \( d_i \) and \( d_j \), merge around a compromise point \( p \). Given the current coalition structure \( D \), the mediator returns a triple \( (d_i, d_j, p) \), where \( p \) is proposed as a new joint position.

Each coalition responds to this suggestion according to a predefined \emph{constitution}, which governs how agents decide whether to join the new coalition. Specifically, agents in \( d_i \) and \( d_j \) vote on whether they prefer the proposed compromise \( p \) over remaining in their current coalition. Based on these votes and the constitution, some agents may transition to the new coalition while others remain.

We first define the voting behavior of an agent before specifying the constitutions that aggregate these votes.

\begin{definition}[Agent, vote]
An \emph{agent} $v$ corresponds to some ideal point $p^v$; and, furthermore, a \emph{vote} of agent $v$ regarding some point $p$ is $vote(v, p) \in \{0, 1\}$ (where $vote(v, p) = 1$ means that $v$ accepts the suggestion to move to a coalition to be formed around $p$).
\end{definition}

Now, a constitution $const$ is defined as follows.

\begin{definition}[Constitution]
A \emph{constitution} $const$ gets as input a tuple $(d_i, d_j, p)$ by the mediator and, when applied on $d_i$ -- and based on the votes of the agents in $d_i$, as described by $\{vote(v, p) : v \in d_i\}$ -- returns an assignment to a coalition for each $v \in d_i$, namely $const(v) \in \{d_i, d^p\}$, where $d^p$ describes the coalition to be possibly-formed around the suggested compromise point $p$.
\end{definition}

Following a suggestion of $(d_i, d_j, p)$ and an application of the constitution $const$ on $d_i$ and $d_j$ (which internally depends on the votes of the agents in both coalitions), the resulting Markov state contains a new coalition structure $D'$ that is defined as follows:\footnote{A coalition with no members can safely be removed from a coalition structure (such that $d'_i$, $d'_j$, and $d^p$ may be empty).}
$D' := D \setminus \{d_i, d_j\} \cup \{d'_i, d'_j, d^p\}, \textrm{where } 
d'_i := (\{v \in d_i : const(v) = d_i\}, p^i)\textrm{; }
d'_j := (\{v \in d_j : const(v) = d_j\}, p^j)\textrm{; }
d^p := (\{v \in d_i \cup d_j : const(v) = d^p\}, p)$.

That is, agents from $d_i$ whom the constitution assigns to $d_i$ remain in it; agents from $d_j$ whom the constitution assigns to $d_j$ remains in it; and agents from $d_i \cup d_j$ whom the constitution assigns to the new coalition around $p$ are being moved there.

\paragraph{A halting condition}
The process halts whenever a coalition that contains an agent majority is being formed; i.e., whenever some $d \in D$, $d = (C, p)$, exists with $\nicefrac{|C|}{|V|} \geq \mathcal{Q}$, where the fraction $\mathcal{Q}\in [0,1]$ can be set to be majority, super majority, or consensus (in our simulation we implement a simple majority).  

\section{Concrete Model Realizations}\label{section:realizations}

We provide concrete realizations of the following ingredients:
agent models (in Section~\ref{section:realization agent}),
coalition constitutions (in Section~\ref{section:realization constitution}),
and mediators (in Section~\ref{section:realization mediator}).
These concrete realizations are used later, for the 2D Euclidean setting presented in the appendix and the setting that involves text documents. Furthermore, some of the details next are needed for the computer-based simulations that follow.~\footnote{We consider realizations of the model in which all agents share the same agent model; all coalitions share the same constitution; and there is one mediator throughout the process. We discuss other options in Section~\ref{section:outlook}.}
%In particular, as we are interested in evaluating the robustness and effectiveness of the model -- and in the evaluation of the effectiveness of different AI-mediators -- we consider different agent models, different coalition constitutions, and different AI-mediators.

% ; note also that this relates to the grassroots aspects discussed in Remark~\ref{remark:grassroots}.

\subsection{One Concrete Agent Model}\label{section:realization agent}

As abstractly stated above, an agent $v$ corresponds to an ideal point $p^v$ and shall have the ability to vote on a proposal $p$ by returning a binary answer in the form of $vote(v, p) \in \{0, 1\}$ -- if $vote(v, p) = 1$, then we say that $v$ \emph{approves} $p$.
Naturally, various realizations of agent models are possible. Below we describe the agent model we use in our theoretical realization (later, in Section~\ref{section:texts} we use a different, LLM-based agent model).
Let us first define a simple, deterministic agent model.

\begin{definition}[A deterministic agent model]
Under the \emph{deterministic agent model}, an agent $v$ within previous coalition $d_i$ with ideal point $p^v$ votes as follows:
$vote(v, p) = 1 \textrm{ if } \mathbf{d}(p^v, p) < \mathbf{d}(p^v, r)$.
\end{definition}

That is, an agent approves a proposal $p$ if $p$ is closer to its ideal point than the status quo $r$, and it disapproves of a proposal $p$ otherwise. Next, as we are interested in modeling agent altruism and flexibility in the process in a naive and intuitive manner (influenced by~\cite{kkk}) we use a probabilistic generalization of the simple model, as described next.

In particular, given the status quo $r$, a proposal $p$, and an agent $v$ with ideal point $p^v$, we define a function $F(r, p, p^v)$ that returns the probability of the agent approving $p$. Specifically, $F(r, p, p^v) \in [0, 1]$. (It may be helpful to note that the deterministic agent model corresponds to the probabilistic model if $F(r, p, p^v) = 1$ whenever $\mathbf{d}(p^v, p) < \mathbf{d}(p^v, r)$.)

Specifically, to model different types of non-deterministic agents, we introduce a parameter $\sigma \geq 0$, where larger values of $\sigma$ results in a more altruistic agent behavior as compared to the simplest agent model described above. Mathematically, we use a half (positive) Gaussian distribution to ``enlarge'' a bit the region for which the agent approves the proposal (i.e., so that an agent will approve a proposal even if it is farther away from its ideal point, compared to the status quo; but with diminishing probability of doing so); formally, we have the following definition of $F$ (note that the \emph{else} case is $0$ in case of $\sigma = 0$):
\[
    F(r, p, p^v) = 
    \begin{cases}
    1, & \text{if }\mathbf{d}(p^v,r) \geq_v \mathbf{d}(p^v,p) \\
    \frac{2}{\sigma_v \sqrt{2\pi}} e^{-\frac{(\mathbf{d}(p^v, p))^2}{2\sigma_v^2}} & \text{else}
    \end{cases}
\]

% Refer to Figure~\ref{Figure A} for an illustration of $F$ for various values of $\sigma$.

\begin{definition}[A probabilistic agent model]
Under the \emph{probabilistic agent model}, an agent $v$ with ideal point $p^v$ votes as follows:
$vote(v, p) = 1 \textrm{ with probability } F(r, p, p^v)$.\footnote{Indeed, for $\sigma = 0$, the probabilistic agent model and the deterministic agent mode coincide.}
\end{definition}

\begin{remark}
   The current agent model assumes that voting behavior depends only on the distance between the proposed point \( p \), the status quo \( r \), and the agent’s ideal point \( p^v \). A natural extension is to allow votes to depend on the anticipated composition of the new coalition. For instance, agents may approve \( p \) only if sufficiently many others are expected to join. 
\end{remark}

\subsection{Two Concrete Constitutions}\label{section:realization constitution}

As abstractly stated above, given a proposal for a coalition \( d_i = (C_i, p_i) \) to move to a new coalition around a compromise point \( p \), a constitution takes the votes of the agents and determines whether any of the coalition members shall move to the new coalition, and, if so, who. We explore two options for such constitutions.

\begin{itemize}

\item 
\textbf{Coalition Discipline:} A new coalition is formed only if at least \( Q \in [0, |C_i|] \) members of \( C_i \) approve the proposal. Formally:
\[
\text{if } |\{v \in C_i : \text{vote}(v, p) = 1\}| \geq Q, \text{ then for each } v \in C_i:
\quad
\text{const}(v) :=
\begin{cases}
d_p & \text{if } \text{vote}(v, p) = 1 \\
d_i & \text{otherwise}
\end{cases}
\]
otherwise, \( \text{const}(v) := d_i \) for all \( v \in C_i \).\footnote{Our model builds upon and generalizes aspects of the framework proposed by Elkind et al. In the case of coalition discipline with deterministic agents and unanimous approval (\( Q = |C_i| \)), we recover their convergence results under the constraint that agents only transition to strictly preferred larger coalitions.\label{footone}}

\item 
\textbf{No Coalition Discipline} is a special case of the above with \( Q = 0 \), where each agent independently decides whether to join the new coalition:
\[
\text{const}(v) :=
\begin{cases}
d_p & \text{if } \text{vote}(v, p) = 1 \\
d_i & \text{otherwise}
\end{cases}
\quad \text{for each } v \in C_i.
\]

\end{itemize}

\begin{remark}
We assume the coalition size \( |C_i| \) is known to its members at the time of voting.
\end{remark}

\begin{remark}
Agent preferences depend only on distance to their ideal point. Coalition discipline imposes coordination constraints but does not affect individual utility.
\end{remark}

\subsection{Several Concrete Mediators}\label{section:realization mediator}

% We move on to consider concrete realizations of the AI-mediators. 
Recall that an mediator takes as input a coalition structure $D$ and returns two coalitions, $d_i$ and $d_j$, and a compromise point $p$.
It is convenient to break the description of our realizations into the two main tasks of mediators, namely: (1) choosing the coalitions $d_i$ and $d_j$ to suggest $p$ to; and (2) choosing the compromise point~$p$ to suggest to $d_i$ and $d_j$.
Note that the role of the AI in our design is in the implementation of such mediators.

\paragraph{Choosing the coalitions $d_i, d_j$}
Our mediators proceed by first computing the centroid of the coalitions' ideal points, weighted by the coalition sizes. Formally:
$centroi\mathbf{d}(D) =\arg\min_{x\in X} \frac{1}{n}\cdot \sum_{i \in |D|} |C_i|\cdot \mathbf{d}(x, p_i)$.
Using the centroid, the mediators consider the distance of each coalition from the centroid, denoted by $\mathbf{d}(p_i, centroid)$. The selection process is guided by a parameter $\alpha \in [-1,1]$, intuitively ranging from whether the closest coalitions to the centroid are preferred ($\alpha=-1$), the furthest ones are preferred ($\alpha=1$), or there is no significance ($\alpha=0$) to the distance from the centroid.

Each coalition $i$ is assigned a score $S_i$ based on its distance from the centroid using the following scoring function $S_i = e^{\alpha\cdot d'(p_i,centroid)}$, 
where $d'(p_i,centroid)\in [0,1]$ is the normalized distance; formally:
\[
d'(p_i,centroid)=\frac{\mathbf{d}(p_i,centroid)}{\arg\max_{j\in |D|}\mathbf{d}(p_j,centroid)}\ .
\]
Subsequently, the mediator assigns a probability $prob(d_i)$ to each coalition $d_i$, proportionate to its scores: $
prob(d_i) = \frac{S_i}{\sum_{i=1}^{|D|}S_i}$.
In practice, the mediator probabilistically chooses one coalition based on these probabilities and then selects the closest coalition to the initially chosen one. 

\paragraph{Choosing the Compromise Point \( p \)}
Given coalitions \( d_i = (C_i, p_i) \) and \( d_j = (C_j, p_j) \), the mediator selects a compromise point \( p \in X \) that minimizes the weighted sum of distances to \( p_i \) and \( p_j \), with weights proportional to coalition sizes:
\[
p = \arg \min_{x \in X} \left( \frac{|C_i|}{|C_i| + |C_j|} \cdot \mathbf{d}(p_i, x) + \frac{|C_j|}{|C_i| + |C_j|} \cdot \mathbf{d}(p_j, x) \right).
\]
In Euclidean space, this reduces to the standard weighted average.

\begin{remark}
The mediator is assumed to know all agents' ideal points. This may result from voluntary disclosure or be treated as a modeling assumption. While this enables computation of globally optimal coalitions under a defined objective, the current mediator applies local, myopic merges. Designing optimal, forward-looking mediators is left for future work.
\end{remark}

% We refer to this mediator as ``the AI-mediator'' since, later, we use a large language model (LLM) to find a text document that maps to the weighted average. However, first, we deliberately chose a simple 2D application to highlight the specifics of our model.

\section{Related Work}

Coalition formation in a metric space has been studied from a multiagent system context~\cite{bulteau2021aggregation,zvi2021iterative,shapiro2022foundations}.
We build upon the theoretical framework of Elkind et al.~\cite{elkind2021united}, which introduced a model for deliberative coalition formation in metric spaces. Their work presents a transition system to capture the dynamics of the coalition formation process. While Elkind et al. describe the formation process in detail, they assume that compromise points are provided by an external source (an oracle), without specifying how these points should be determined. Our contribution addresses this gap by introducing \emph{mediators} that algorithmically suggest compromise points, allowing coalitions to unite around majority-supported proposals. We implement and optimize these mediators to make the coalition formation process both practical and efficient. We also demonstrate that, under a specific configuration of our model, it aligns with Elkind's model, thereby inheriting their theoretical results for that configuration thus our model generalizes Elkind's model. For the general case we show simulations that show convergence rates are very good even for large instances.
In developing mediators, we utilize NLP techniques and LLMs; this relates to NLP-based recommendation systems~\cite{balush2021recommendation,wang2016compare}, where models suggest content based on user preferences, and to recent work in \emph{Generative Social Choice}~\cite{fish2023generative}, which explores the use of LLMs to generate representative statements for social choice tasks. However, our work differs by focusing on identifying compromise points in the coalition formation process, where the goal is to find a majority-supported text or proposal.
Another relevant line of work is by Bakker et al.~\cite{bakker2022fine}, who study how machines can assist in finding agreements among individuals with diverse preferences. Their approach fine-tunes LLMs to generate statements that maximize the expected approval of a group, which is conceptually similar to our use of AI for proposing compromise points. However, our model incorporates an iterative coalition formation process, making it distinct in its operational dynamics. We also mention Yang et al.~\cite{yang2024llm} that investigate how GPT-4 and LLaMA-2 behave in voting scenarios compared to human voters. They show that voting methods, presentation order, and temperature settings can significantly influence LLM choices, often reducing preference diversity and risking bias. 

In the context of \emph{Dynamic Coalition Formation}, there is significant prior work on how agents with diverse preferences form and adapt coalitions to achieve consensus~\cite{shehory1998multi,KONISHI20031}. This is relevant, as coalition formation plays a key role in decision-making processes, especially when agents aim to form majority-supported agreements~\cite{rahwan2015coalition}. Our approach to coalition formation in metric spaces also draws on existing research in spatial voting models~\cite{enelow1984spatial,heitzig2024fair}.
 We are also motivated by psychological research on the ability of agents to objectively evaluate proposals. Mikhaylovskaya et al.~\cite{mikhaylovskaya2024building} provide evidence that AI-based mediators can mitigate human biases, making AI a promising tool for generating neutral, data-driven compromise points. This motivates our use of AI-mediated coalition formation, where agents can evaluate AI-suggested compromise points to find collective agreements.
Our work also draws inspiration from negotiation-based approaches to coalition formation~\cite{rahwan2007algorithms,haynes2020introduction,sarkar2022survey,elkind2021united,janovsky2016multi,wanyama2006negotiation,beer1999negotiation,espinasse1997negotiation}, which offer valuable insights into how agents with divergent preferences negotiate and form coalitions. These approaches further reinforce the relevance of AI based mediation in improving the efficiency and effectiveness of coalition formation in multi-agent systems.
\section{Mediators in a Textual Space}\label{section:texts}

As our ultimate goal relates to text aggregation -- i.e., to enable an agent community to converge towards a majority-supported textual document. So, we wish to utilize the mediators framework  (demonstrated also in a Euclidean space in the appendix) to a setting in which the metric space contains textual documents and coalitions form around different texts, until a majority-supported textual document is identified.
We describe our specific model; and then report on computer-based simulations.

\subsection{Modeling Mediators in a Textual Space}

Our general solution works as follows.
We use word embedding (a standard NLP technique) to translate texts into numerical-valued vectors; this is crucial as, after applying such a word embedding, we are then able to compute distances between the embedded coalition ideal points and use the mediators of the Euclidean space. In this work, we use Google's Universal Sentence Encoder~\cite{cer2018universal}: this is a pre-trained model that converts sentences into fixed-size vectors, capturing their semantic meanings. (The Universal Sentence Encoder is designed to generate 512-dimensional embedding vectors, providing a semantic representation of sentences.)
Thus our embedded metric space contains as elements all those $512$-length vectors that can be the output of the Universal Sentence Encoder. As a distance measure in this space, we use the square root of the cosine-based dissimilarity~\cite{schubert2019elki}, a commonly used pseudo metric in NLP.%
\footnote{Formally, we define
\[
d_{\text{cos}}(A, B) := \sqrt{2 - 2 \cdot \frac{A \cdot B}{\|A\| \cdot \|B\|}} \in [0, 2],
\]
which corresponds to \( \sqrt{2} \cdot \sin(\theta/2) \), where \( \theta \) is the angle between vectors \( A \) and \( B \). This is a metric when restricted to the unit sphere (or more generally, to rays through the origin), but only a pseudo metric over the entire embedding space: co-linear vectors have distance zero, even if they differ in magnitude. Since our mediator constructs compromise vectors via (non-normalized) weighted averages, we do not assume that vectors lie on a sphere.}
The mediator guides the coalition formation by proposing sentences to two coalitions within the given word limit (in our simulations, $15$ words). In each iteration, the mediator's goal is to find two coalitions to suggest a sentence that minimizes the \emph{squared cosine similarity} pseudo metric between the embedding vector of the chosen sentence and the weighted average of the embedding vectors of the two coalition points.

Our approach to coalition formation in the domain of text relies on the integration of OpenAI's GPT-3.5-turbo-1106 model (\url{https://openai.com/blog/chatgpt}) with a temperature parameter (responsible for the randomness of results)  of $0.75$ as recommended in some of the documentation to provide a good trade-off for applications like ours where the output should be coherent but still allow for some diversity and creativity. This LLM takes $3$ key roles within our framework:
\begin{enumerate}

\item It generates sentences that act as agent ideal points.

\item It constructs initial singleton coalitions mirroring these ideal sentences if the coalition formation process introduces noise (i.e., for simulations runs with $I = true$).

\item It proposes diverse options for aggregating two sentences, presenting methods to combine opinions from different coalitions represented as text. The process then determines the most suitable sentence by evaluating which has the embedding that is the closest to the weighted average of the two coalition embedded sentences.

\end{enumerate}

\begin{remark}
We assume that Euclidean distance in the embedding space reflects agent preferences. That is, texts closer to an agent's embedded ideal point are considered more preferable. This assumption connects the embedding to the distance-based agent model, though it may not hold uniformly across domains.
\end{remark}

\subsection{Simulations-Based Analysis}

We conducted simulations to assess the robustness and resilience of the model; done as follows:
\begin{itemize}

\item \textbf{Parameter Tests and Scale:} The simulations were conducted with different numbers of agents, specifically \( n \in \{10, 20, 30, 40, 50, 100, 1000\} \). We varied the parameters \(\sigma \in \{0,1,1.5\}\) and \(\alpha \in \{-1,0,1\}\), while also setting the boolean variable \(C\)--- enforcing coalition discipline or not. Each parameter combination was tested across 50 repetitions, with all ideal sentences generated sharing a predetermined topic of ways to address global warming.

\item \textbf{Coalition Formation Process:}
We employed an iterative pursuit in an embedding space using \emph{squared cosine similarity} pseudo metric;
% %    
%     \item Unlike the location usecase, there was no exploration of different Gaussian Mixture Model (GMM) distributions for ideal points, as they were generated by GPT-3.5 (i.e., parameter $g$ was not included here). 
%
the prompt given to GPT was: ``Give me T different sentences that are well structured about how to deal with Y with at most of 15 words'' (T being the number of agents, and Y being any topic -- global warming in our case);
to initialize the singleton coalition with introducing noise ($I=True$), the LLM was requested to provide a sentence resembling the ideal sentence of each agent, rather than introducing additional noise through a normal distribution as conducted in the euclidean case presented in the appendix. These function as the singleton coalition sentences to be embedded into the Euclidean space. The prompt given to the GPT was: ``Give me a well-structured sentence with a maximum of 15 words, resembling this sentence: Z'' (where Z represents an ideal sentence of an agent).
    
\item \textbf{Sentence Selection Process:}
For each proposed sentence to the two coalitions, the LLM was tasked with generating $10$ sentences that effectively combined both coalition sentences. We followed best practices for structured prompt design and multi-step reasoning~\cite{kojima2022large}, including these concepts:
\begin{itemize}
    \item \textit{Structured Prompt Design:}  Prompts should provide clear and concise instructions, ensuring that the LLM produced well-structured sentences.
    
    \item \textit{Encouraging Multi-Step Reasoning:} Prompts should be designed to guide the LLM through step-by-step reasoning, leveraging Zero-shot Chain of Thought (CoT) techniques to handle the task effectively.
    
    \item \textit{No Task-Specific Examples Needed:} Prompts should avoid the need for specific examples or task-specific training, enabling the model to generalize across different tasks.
    
\end{itemize}
    
We used the following prompts and messages given to GPT 3.5 ($5$ options in total):
\begin{itemize}

\item
\underline{Mediator 1}:
\textbf{Prompt:} Generate $10$ possible different well-structured sentences that aggregate the following two sentences. Make sure each sentence has at most 15 words. Number your answers (i.e., 1), 2), 3), 4), 5), and so on) for each sentence you propose.
\textbf{Message:} You are a mediator trying to find agreed wording for how to deal with global warming based on existing sentences. Give a straightforward answer with no introduction to help people reach an agreed wording of a coherent sentence.
(The proposed sentence was selected based on the minimal \emph{squared cosine similarity} pseudo metric between its embedding and the weighted average embedding vector, considering the two embedding vectors of the coalitions and their sizes.)

\item
\underline{Mediator 2}:
\textbf{Prompt:} Generate 10 concise and clear sentences that blend the following two sentences into one coherent idea:
Ensure each sentence is no longer than 15 words. Number your answers (i.e., 1), 2), 3), 4), 5), and so on) for each sentence you propose.
\textbf{Message:} As a mediator, you need to find a consensus on global warming solutions. Provide straightforward and numbered suggestions to help reach a clear and agreed-upon sentence.

\item
\underline{Mediator 3}:
\textbf{Prompt:} Create 10 unique, well-structured sentences that combine these two sentences into one unified thought:
Each sentence should be a maximum of 15 words. Number your answers (i.e., 1), 2), 3), 4), 5), and so on) for each sentence you propose.
\textbf{Message:} You are acting as a mediator to achieve a common statement on global warming. Give direct and numbered suggestions to assist in forming a unified and coherent sentence.

\item
\underline{Mediator 4}:
This baseline mediator involved soliciting several possibilities for sentence aggregation from the GPT and then selecting the sentence that minimized the distance from the average embedding vector of the two coalitions. Instead, we simply requested GPT to provide a single sentence. The prompt and message given to GPT were the same as those given for Option 1, but instead of 10 sentences, it was asked for 1 sentence only. 

\item
\underline{Mediator 5}:
This second baseline mediator denoted by \textbf{Option 5}, was to ask GPT for a completely random sentence.

\end{itemize}

\end{itemize}

\begin{table}[t]
    \centering
    \begin{tabular}{cc}
        \toprule
        Option & Mean Number of Iterations \\
        \midrule
        Option 1 & 4.8000 \\
        Option 2 & 5.0750 \\
        Option 3 & 5.5250 \\
        Option 4 & 7.5000 \\
        Option 5 & 41.8125 \\
        \bottomrule
    \end{tabular}
    \caption{A comparison of different mediators-each corresponding to different prompts and LLM-usage strategy. The mean number of iterations until convergence is shown, validated with $95\%$ confidence using ANOVA and Tukey HSD.}\label{tuki}
\end{table}
We tested the number of iterations needed for coalitions to converge on a compromise, and the average distance between the compromise document and the ideal document of each agent within the coalition that halted the process.

\section{Outlook and Discussion}\label{section:outlook}

Our findings closely align with the Euclidean case presented in the appendix:  

1) Processes with coalition discipline and deterministic agents always exhibit some cases of non-convergence (defined as exceeding 10,000 iterations), whereas all other combinations result in convergence;  
2) A higher number of agents leads to more iterations;  
3) Increasing $\alpha$ enlarges the mean average distance between the compromise sentence of the largest coalition and the ideal sentences of its members;  
4) Coalition discipline reduces this mean average distance.  

We also analyze the performance of different mediators, summarized in Table~\ref{tuki}. An ANOVA test confirms statistically significant differences in the number of iterations across mediation approaches. A post-hoc Tukey HSD test (see appendix) further identifies significant pairwise differences, revealing that Option 1 achieves the fewest iterations on average.  

\begin{remark}  
We conducted additional experiments using GPT-3 Davinci and GPT-4o Mini, both with a temperature of 0.75. As their results followed the same patterns and led to identical conclusions, we omit them here for brevity.  
\end{remark}  

\subsection{Interpretation}  

Our simulations demonstrate the effectiveness of \emph{AI-mediated coalition formation}, particularly when leveraging \emph{Large Language Models (LLMs)}. AI-mediation significantly reduces the number of iterations required for coalitions to reach a compromise while minimizing the average distance between the final compromise and individual agents' ideal documents. Notably, LLMs combined with distance-based optimization consistently accelerate convergence compared to simpler mediator approaches.  

The statistical tests reinforce that meaningful differences exist among mediation strategies, underscoring the adaptability of AI-mediation to different scenarios. The superior performance of Option 2 in minimizing iterations further suggests that careful tuning of the mediator’s behavior can yield substantial efficiency gains.

\subsection{Future Work}

Several directions for future research are outlined below:
\begin{itemize}

\item \textbf{Theoretical Guarantees:} Analyze convergence under relaxed rationality and probabilistic behavior; study stability and fairness properties.

\item \textbf{Strategic Behavior:} Extend the model to a game-theoretic setting that accounts for strategic agents who may misreport preferences and anticipate coalition dynamics.

\item \textbf{Scalability:} Develop efficient mediator selection, distributed implementations, and hierarchical coalition structures for large-scale settings.

\item \textbf{Bias and Interpretability:} Address AI-induced bias, enforce fairness constraints, and improve mediator transparency.

\item \textbf{Application Domains:} Apply the model to other contexts, such as participatory budgeting, resource allocation, and collaborative drafting.

\item \textbf{Empirical Evaluation:} Test the framework in real-world environments (e.g., DAOs, Wikipedia); assess adoption via human studies.

\item \textbf{Adaptive Mediators:} Use reinforcement learning or game-theoretic tools to adapt mediator strategies over time.

\item \textbf{Decentralized Use:} Support coalition-local mediator usage in decentralized systems with autonomous agent groups.

\item \textbf{Proportionality:} Mitigate majority dominance using methods like Phragmén’s rule~\cite{peters2024proportional} to ensure proportional influence in aggregation.

\item \textbf{Forward-Looking Planning:} Extend mediators to evaluate merge sequences that optimize objectives (e.g., minimum distance or maximal support), under different assumptions about agent information.
\item \textbf{Deliberation and Communication:} Extend the model to allow agent-to-agent communication, enabling persuasion or belief updates during the process.

\item \textbf{Context-Sensitive Voting Behavior:} Our model compares proposals to a fixed status quo, following Elkind et al.~\cite{elkind2021united}. A natural extension is to compare proposals to the current coalition point \( p_c \), i.e., accept \( p \) if \( d(p^v, p) < d(p^v, p_c) \). A more refined model would also consider coalition size and composition—agents may prefer large coalitions for influence, or avoid those with ideologically distant members.

\end{itemize}

\paragraph{Acknowledgment. }
Authors Briman and Talmon acknowledge support by the European Union under the Horizon Europe project \href{https://perycles-project.eu/}{Perycles} (Participatory Democracy that Scales). Views and opinions expressed are however those of the author(s) only and do not necessarily reflect those of the European Union or the European Research Executive Agency. Neither the European Union nor the granting authority can be held responsible for them.
\smallskip
\begin{center}
\includegraphics[width=0.5\textwidth]{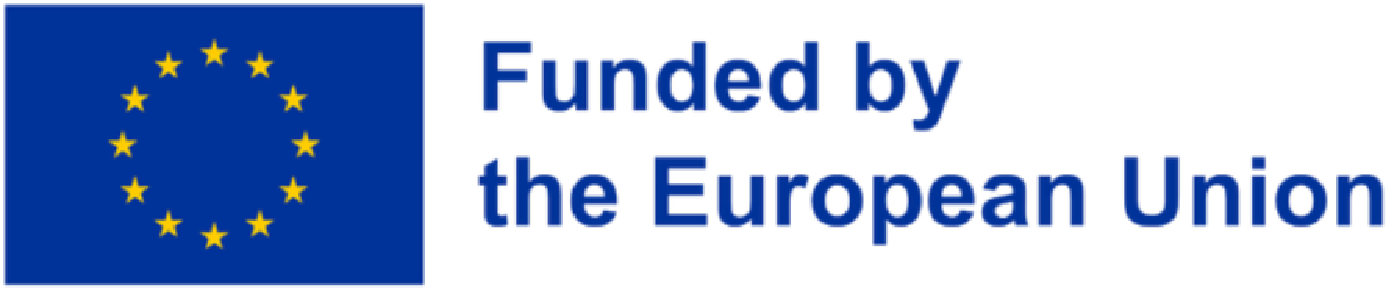}
\end{center}

\bibliographystyle{eptcs}
\bibliography{TARK2025/bib}

\begin{thebibliography}{10}
\providecommand{\bibitemdeclare}[2]{}
\providecommand{\surnamestart}{}
\providecommand{\surnameend}{}
\providecommand{\urlprefix}{Available at }
\providecommand{\url}[1]{\texttt{#1}}
\providecommand{\href}[2]{\texttt{#2}}
\providecommand{\urlalt}[2]{\href{#1}{#2}}
\providecommand{\doi}[1]{doi:\urlalt{https://doi.org/#1}{#1}}
\providecommand{\eprint}[1]{arXiv:\urlalt{https://arxiv.org/abs/#1}{#1}}
\providecommand{\bibinfo}[2]{#2}

\bibitemdeclare{article}{bakker2022fine}
\bibitem{bakker2022fine}
\bibinfo{author}{Michiel \surnamestart Bakker\surnameend}, \bibinfo{author}{Martin \surnamestart Chadwick\surnameend}, \bibinfo{author}{Hannah \surnamestart Sheahan\surnameend}, \bibinfo{author}{Michael \surnamestart Tessler\surnameend}, \bibinfo{author}{Lucy \surnamestart Campbell-Gillingham\surnameend}, \bibinfo{author}{Jan \surnamestart Balaguer\surnameend}, \bibinfo{author}{Nat \surnamestart McAleese\surnameend}, \bibinfo{author}{Amelia \surnamestart Glaese\surnameend}, \bibinfo{author}{John \surnamestart Aslanides\surnameend}, \bibinfo{author}{Matt \surnamestart Botvinick\surnameend} et~al. (\bibinfo{year}{2022}): \emph{\bibinfo{title}{Fine-tuning language models to find agreement among humans with diverse preferences}}.
\newblock {\slshape \bibinfo{journal}{Advances in Neural Information Processing Systems}} \bibinfo{volume}{35}, pp. \bibinfo{pages}{38176--38189}, \doi{10.48550/arXiv.2211.15006}.

\bibitemdeclare{misc}{balush2021recommendation}
\bibitem{balush2021recommendation}
\bibinfo{author}{Illia \surnamestart Balush\surnameend}, \bibinfo{author}{Victoria \surnamestart Vysotska\surnameend} \& \bibinfo{author}{Solomiia \surnamestart Albota\surnameend} (\bibinfo{year}{2021}): \emph{\bibinfo{title}{Recommendation System Development Based on Intelligent Search, NLP and Machine Learning Methods.}}

\bibitemdeclare{article}{beer1999negotiation}
\bibitem{beer1999negotiation}
\bibinfo{author}{Martin \surnamestart Beer\surnameend}, \bibinfo{author}{Mark \surnamestart d'Inverno\surnameend}, \bibinfo{author}{Michael \surnamestart Luck\surnameend}, \bibinfo{author}{Nick \surnamestart Jennings\surnameend}, \bibinfo{author}{Chris \surnamestart Preist\surnameend} \& \bibinfo{author}{Michael \surnamestart Schroeder\surnameend} (\bibinfo{year}{1999}): \emph{\bibinfo{title}{Negotiation in multi-agent systems}}.
\newblock {\slshape \bibinfo{journal}{The Knowledge Engineering Review}} \bibinfo{volume}{14}(\bibinfo{number}{3}), pp. \bibinfo{pages}{285--289}, \doi{10.1017/S0269888999003021}.

\bibitemdeclare{article}{bogomolnaia2007euclidean}
\bibitem{bogomolnaia2007euclidean}
\bibinfo{author}{Anna \surnamestart Bogomolnaia\surnameend} \& \bibinfo{author}{Jean-Fran{\c{c}}ois \surnamestart Laslier\surnameend} (\bibinfo{year}{2007}): \emph{\bibinfo{title}{Euclidean preferences}}.
\newblock {\slshape \bibinfo{journal}{Journal of Mathematical Economics}} \bibinfo{volume}{43}(\bibinfo{number}{2}), pp. \bibinfo{pages}{87--98}, \doi{10.1016/j.jmateco.2006.09.004}.

\bibitemdeclare{article}{bulteau2021aggregation}
\bibitem{bulteau2021aggregation}
\bibinfo{author}{Laurent \surnamestart Bulteau\surnameend}, \bibinfo{author}{Gal \surnamestart Shahaf\surnameend}, \bibinfo{author}{Ehud \surnamestart Shapiro\surnameend} \& \bibinfo{author}{Nimrod \surnamestart Talmon\surnameend} (\bibinfo{year}{2021}): \emph{\bibinfo{title}{Aggregation over metric spaces: Proposing and voting in elections, budgeting, and legislation}}.
\newblock {\slshape \bibinfo{journal}{Journal of Artificial Intelligence Research}} \bibinfo{volume}{70}, pp. \bibinfo{pages}{1413--1439}, \doi{10.1613/jair.1.12388}.

\bibitemdeclare{article}{cer2018universal}
\bibitem{cer2018universal}
\bibinfo{author}{Daniel \surnamestart Cer\surnameend}, \bibinfo{author}{Yinfei \surnamestart Yang\surnameend}, \bibinfo{author}{Sheng-yi \surnamestart Kong\surnameend}, \bibinfo{author}{Nan \surnamestart Hua\surnameend}, \bibinfo{author}{Nicole \surnamestart Limtiaco\surnameend}, \bibinfo{author}{Rhomni~St \surnamestart John\surnameend}, \bibinfo{author}{Noah \surnamestart Constant\surnameend}, \bibinfo{author}{Mario \surnamestart Guajardo-Cespedes\surnameend}, \bibinfo{author}{Steve \surnamestart Yuan\surnameend}, \bibinfo{author}{Chris \surnamestart Tar\surnameend} et~al. (\bibinfo{year}{2018}): \emph{\bibinfo{title}{Universal sentence encoder}}.
\newblock {\slshape \bibinfo{journal}{arXiv preprint arXiv:1803.11175}} \bibinfo{volume}{N/A}(\bibinfo{number}{N/A}), p. \bibinfo{pages}{N/A}, \doi{10.18653/v1/D18-2029}.

\bibitemdeclare{misc}{nlp}
\bibitem{nlp}
\bibinfo{author}{K.~R. \surnamestart Chowdhary\surnameend} (\bibinfo{year}{2020}): \emph{\bibinfo{title}{Natural language processing for word sense disambiguation and information extraction}}, \doi{10.48550/arXiv.2004.02256}.

\bibitemdeclare{inproceedings}{elkind2017multiwinner}
\bibitem{elkind2017multiwinner}
\bibinfo{author}{Edith \surnamestart Elkind\surnameend}, \bibinfo{author}{Piotr \surnamestart Faliszewski\surnameend}, \bibinfo{author}{Jean-Fran{\c{c}}ois \surnamestart Laslier\surnameend}, \bibinfo{author}{Piotr \surnamestart Skowron\surnameend}, \bibinfo{author}{Arkadii \surnamestart Slinko\surnameend} \& \bibinfo{author}{Nimrod \surnamestart Talmon\surnameend} (\bibinfo{year}{2017}): \emph{\bibinfo{title}{What do multiwinner voting rules do? An experiment over the two-dimensional euclidean domain}}.
\newblock In: {\slshape \bibinfo{booktitle}{Proceedings of the AAAI Conference on Artificial Intelligence}}, \bibinfo{volume}{31}, \bibinfo{publisher}{AAAI Press}, \bibinfo{address}{Palo Alto, CA}, p. \bibinfo{pages}{N/A}, \doi{10.1609/aaai.v31i1.10612}.

\bibitemdeclare{inproceedings}{elkind2021united}
\bibitem{elkind2021united}
\bibinfo{author}{Edith \surnamestart Elkind\surnameend}, \bibinfo{author}{Davide \surnamestart Grossi\surnameend}, \bibinfo{author}{Ehud \surnamestart Shapiro\surnameend} \& \bibinfo{author}{Nimrod \surnamestart Talmon\surnameend} (\bibinfo{year}{2021}): \emph{\bibinfo{title}{United for change: deliberative coalition formation to change the status quo}}.
\newblock In: {\slshape \bibinfo{booktitle}{Proceedings of the AAAI Conference on Artificial Intelligence}}, \bibinfo{volume}{35}, \bibinfo{publisher}{AAAI Press}, \bibinfo{address}{Palo Alto, CA}, pp. \bibinfo{pages}{5339--5346}, \doi{10.1609/aaai.v35i6.16673}.

\bibitemdeclare{book}{enelow1984spatial}
\bibitem{enelow1984spatial}
\bibinfo{author}{James~M \surnamestart Enelow\surnameend} \& \bibinfo{author}{Melvin~J \surnamestart Hinich\surnameend} (\bibinfo{year}{1984}): \emph{\bibinfo{title}{The spatial theory of voting: An introduction}}.
\newblock \bibinfo{publisher}{Cambridge University Press}, \bibinfo{address}{Cambridge, UK}.

\bibitemdeclare{article}{espinasse1997negotiation}
\bibitem{espinasse1997negotiation}
\bibinfo{author}{Bernard \surnamestart Espinasse\surnameend}, \bibinfo{author}{Guy \surnamestart Picolet\surnameend} \& \bibinfo{author}{Eugene \surnamestart Chouraqui\surnameend} (\bibinfo{year}{1997}): \emph{\bibinfo{title}{Negotiation support systems: A multi-criteria and multi-agent approach}}.
\newblock {\slshape \bibinfo{journal}{European Journal of Operational Research}} \bibinfo{volume}{103}(\bibinfo{number}{2}), pp. \bibinfo{pages}{389--409}, \doi{10.1016/S0377-2217(97)00127-6}.

\bibitemdeclare{article}{fish2023generative}
\bibitem{fish2023generative}
\bibinfo{author}{Sara \surnamestart Fish\surnameend}, \bibinfo{author}{Paul \surnamestart G{\"o}lz\surnameend}, \bibinfo{author}{David~C \surnamestart Parkes\surnameend}, \bibinfo{author}{Ariel~D \surnamestart Procaccia\surnameend}, \bibinfo{author}{Gili \surnamestart Rusak\surnameend}, \bibinfo{author}{Itai \surnamestart Shapira\surnameend} \& \bibinfo{author}{Manuel \surnamestart W{\"u}thrich\surnameend} (\bibinfo{year}{2023}): \emph{\bibinfo{title}{Generative Social Choice}}.
\newblock {\slshape \bibinfo{journal}{arXiv preprint arXiv:2309.01291}} \bibinfo{volume}{N/A}(\bibinfo{number}{N/A}), p. \bibinfo{pages}{N/A}, \doi{10.48550/arXiv.2309.01291}.

\bibitemdeclare{misc}{dao}
\bibitem{dao}
\bibinfo{author}{Samer \surnamestart Hassan\surnameend} \& \bibinfo{author}{Primavera \surnamestart De~Filippi\surnameend} (\bibinfo{year}{2021}): \emph{\bibinfo{title}{Decentralized autonomous organization}}.

\bibitemdeclare{article}{haynes2020introduction}
\bibitem{haynes2020introduction}
\bibinfo{author}{Teresa~W \surnamestart Haynes\surnameend}, \bibinfo{author}{Jason~T \surnamestart Hedetniemi\surnameend}, \bibinfo{author}{Stephen~T \surnamestart Hedetniemi\surnameend}, \bibinfo{author}{Alice~A \surnamestart McRae\surnameend} \& \bibinfo{author}{Raghuveer \surnamestart Mohan\surnameend} (\bibinfo{year}{2020}): \emph{\bibinfo{title}{Introduction to coalitions in graphs}}.
\newblock {\slshape \bibinfo{journal}{AKCE International Journal of Graphs and Combinatorics}} \bibinfo{volume}{17}(\bibinfo{number}{2}), pp. \bibinfo{pages}{653--659}, \doi{10.1080/09728600.2020.1832874}.

\bibitemdeclare{article}{heitzig2024fair}
\bibitem{heitzig2024fair}
\bibinfo{author}{Jobst \surnamestart Heitzig\surnameend}, \bibinfo{author}{Forest~W \surnamestart Simmons\surnameend} \& \bibinfo{author}{Sara~M \surnamestart Constantino\surnameend} (\bibinfo{year}{2024}): \emph{\bibinfo{title}{Fair group decisions via non-deterministic proportional consensus}}.
\newblock {\slshape \bibinfo{journal}{Social Choice and Welfare}} \bibinfo{volume}{N/A}(\bibinfo{number}{N/A}), pp. \bibinfo{pages}{1--27}, \doi{10.1007/s00355-024-01524-3}.

\bibitemdeclare{inproceedings}{janovsky2016multi}
\bibitem{janovsky2016multi}
\bibinfo{author}{Pavel \surnamestart Janovsky\surnameend} \& \bibinfo{author}{Scott~A \surnamestart DeLoach\surnameend} (\bibinfo{year}{2016}): \emph{\bibinfo{title}{Multi-agent simulation framework for large-scale coalition formation}}.
\newblock In: {\slshape \bibinfo{booktitle}{2016 IEEE/WIC/ACM International Conference on Web Intelligence (WI)}}, \bibinfo{publisher}{IEEE}, \bibinfo{address}{N/A}, pp. \bibinfo{pages}{343--350}, \doi{10.1109/WI.2016.0055}.

\bibitemdeclare{inproceedings}{kkk}
\bibitem{kkk}
\bibinfo{author}{Anna~Maria \surnamestart Kerkmann\surnameend} (\bibinfo{year}{2022}): \emph{\bibinfo{title}{Stability, Fairness, and Altruism in Coalition Formation}}.
\newblock In \bibinfo{editor}{Dorothea \surnamestart Baumeister\surnameend} \& \bibinfo{editor}{J{\"o}rg \surnamestart Rothe\surnameend}, editors: {\slshape \bibinfo{booktitle}{Multi-Agent Systems}}, \bibinfo{publisher}{Springer International Publishing}, \bibinfo{address}{Cham}, pp. \bibinfo{pages}{427--430}, \doi{10.1007/978-3-031-20614-6_25}.

\bibitemdeclare{article}{kojima2022large}
\bibitem{kojima2022large}
\bibinfo{author}{Takeshi \surnamestart Kojima\surnameend}, \bibinfo{author}{Shixiang~Shane \surnamestart Gu\surnameend}, \bibinfo{author}{Machel \surnamestart Reid\surnameend}, \bibinfo{author}{Yutaka \surnamestart Matsuo\surnameend} \& \bibinfo{author}{Yusuke \surnamestart Iwasawa\surnameend} (\bibinfo{year}{2022}): \emph{\bibinfo{title}{Large language models are zero-shot reasoners}}.
\newblock {\slshape \bibinfo{journal}{Advances in Neural Information Processing Systems}} \bibinfo{volume}{35}, pp. \bibinfo{pages}{22199--22213}, \doi{10.48550/arXiv.2205.11916}.

\bibitemdeclare{article}{KONISHI20031}
\bibitem{KONISHI20031}
\bibinfo{author}{Hideo \surnamestart Konishi\surnameend} \& \bibinfo{author}{Debraj \surnamestart Ray\surnameend} (\bibinfo{year}{2003}): \emph{\bibinfo{title}{Coalition formation as a dynamic process}}.
\newblock {\slshape \bibinfo{journal}{Journal of Economic Theory}} \bibinfo{volume}{110}(\bibinfo{number}{1}), pp. \bibinfo{pages}{1--41}, \doi{10.1016/S0022-0531(03)00004-8}.

\bibitemdeclare{article}{kraus1997negotiation}
\bibitem{kraus1997negotiation}
\bibinfo{author}{Sarit \surnamestart Kraus\surnameend} (\bibinfo{year}{1997}): \emph{\bibinfo{title}{Negotiation and cooperation in multi-agent environments}}.
\newblock {\slshape \bibinfo{journal}{Artificial Intelligence}} \bibinfo{volume}{94}(\bibinfo{number}{1--2}), pp. \bibinfo{pages}{79--97}, \doi{10.1016/S0004-3702(97)00025-8}.

\bibitemdeclare{misc}{van2008visualizing}
\bibitem{van2008visualizing}
\bibinfo{author}{Laurens \surnamestart Van~der Maaten\surnameend} \& \bibinfo{author}{Geoffrey \surnamestart Hinton\surnameend} (\bibinfo{year}{2008}): \emph{\bibinfo{title}{Visualizing data using t-SNE}}.

\bibitemdeclare{article}{maturana1996multi}
\bibitem{maturana1996multi}
\bibinfo{author}{Francisco~P \surnamestart Maturana\surnameend} \& \bibinfo{author}{Douglas~H \surnamestart Norrie\surnameend} (\bibinfo{year}{1996}): \emph{\bibinfo{title}{Multi-agent mediator architecture for distributed manufacturing}}.
\newblock {\slshape \bibinfo{journal}{Journal of Intelligent Manufacturing}} \bibinfo{volume}{7}, pp. \bibinfo{pages}{257--270}, \doi{10.1007/BF00124828}.

\bibitemdeclare{article}{mikhaylovskaya2024building}
\bibitem{mikhaylovskaya2024building}
\bibinfo{author}{Anna \surnamestart Mikhaylovskaya\surnameend} \& \bibinfo{author}{{\'E}lise \surnamestart Roum{\'e}as\surnameend} (\bibinfo{year}{2024}): \emph{\bibinfo{title}{Building trust with digital democratic innovations}}.
\newblock {\slshape \bibinfo{journal}{Ethics and Information Technology}} \bibinfo{volume}{26}(\bibinfo{number}{1}), p.~\bibinfo{pages}{1}, \doi{10.1007/s10676-023-09736-4}.

\bibitemdeclare{incollection}{peters2024proportional}
\bibitem{peters2024proportional}
\bibinfo{author}{Dominik \surnamestart Peters\surnameend} (\bibinfo{year}{2024}): \emph{\bibinfo{title}{Proportional Representation for Artificial Intelligence}}.
\newblock In: {\slshape \bibinfo{booktitle}{ECAI 2024}}, \bibinfo{publisher}{IOS Press}, pp. \bibinfo{pages}{27--31}, \doi{10.3233/FAIA240463}.

\bibitemdeclare{phdthesis}{rahwan2007algorithms}
\bibitem{rahwan2007algorithms}
\bibinfo{author}{Talal \surnamestart Rahwan\surnameend} (\bibinfo{year}{2007}): \emph{\bibinfo{title}{Algorithms for coalition formation in multi-agent systems}}.
\newblock Ph.D. thesis, \bibinfo{school}{University of Southampton}, \bibinfo{address}{N/A}.

\bibitemdeclare{article}{rahwan2015coalition}
\bibitem{rahwan2015coalition}
\bibinfo{author}{Talal \surnamestart Rahwan\surnameend}, \bibinfo{author}{Tomasz~P \surnamestart Michalak\surnameend}, \bibinfo{author}{Michael \surnamestart Wooldridge\surnameend} \& \bibinfo{author}{Nicholas~R \surnamestart Jennings\surnameend} (\bibinfo{year}{2015}): \emph{\bibinfo{title}{Coalition structure generation: A survey}}.
\newblock {\slshape \bibinfo{journal}{Artificial Intelligence}} \bibinfo{volume}{229}, pp. \bibinfo{pages}{139--174}, \doi{10.1016/j.artint.2015.08.004}.

\bibitemdeclare{incollection}{rosenfeld2016strategical}
\bibitem{rosenfeld2016strategical}
\bibinfo{author}{Ariel \surnamestart Rosenfeld\surnameend} \& \bibinfo{author}{Sarit \surnamestart Kraus\surnameend} (\bibinfo{year}{2016}): \emph{\bibinfo{title}{Strategical argumentative agent for human persuasion}}.
\newblock In: {\slshape \bibinfo{booktitle}{ECAI 2016}}, \bibinfo{publisher}{IOS Press}, pp. \bibinfo{pages}{320--328}, \doi{10.3233/978-1-61499-672-9-320}.

\bibitemdeclare{article}{sarkar2022survey}
\bibitem{sarkar2022survey}
\bibinfo{author}{Samriddhi \surnamestart Sarkar\surnameend}, \bibinfo{author}{Mariana \surnamestart Curado~Malta\surnameend} \& \bibinfo{author}{Animesh \surnamestart Dutta\surnameend} (\bibinfo{year}{2022}): \emph{\bibinfo{title}{A survey on applications of coalition formation in multi-agent systems}}.
\newblock {\slshape \bibinfo{journal}{Concurrency and Computation: Practice and Experience}} \bibinfo{volume}{34}(\bibinfo{number}{11}), p. \bibinfo{pages}{e6876}, \doi{10.1002/cpe.6876}.

\bibitemdeclare{article}{schubert2019elki}
\bibitem{schubert2019elki}
\bibinfo{author}{Erich \surnamestart Schubert\surnameend} \& \bibinfo{author}{Arthur \surnamestart Zimek\surnameend} (\bibinfo{year}{2019}): \emph{\bibinfo{title}{ELKI: A large open-source library for data analysis-ELKI Release 0.7.5 "Heidelberg"}}.
\newblock {\slshape \bibinfo{journal}{arXiv preprint arXiv:1902.03616}} \bibinfo{volume}{N/A}(\bibinfo{number}{N/A}), p. \bibinfo{pages}{N/A}, \doi{10.48550/arXiv.1902.03616}.

\bibitemdeclare{article}{shapiro2022foundations}
\bibitem{shapiro2022foundations}
\bibinfo{author}{Ehud \surnamestart Shapiro\surnameend} \& \bibinfo{author}{Nimrod \surnamestart Talmon\surnameend} (\bibinfo{year}{2022}): \emph{\bibinfo{title}{Foundations for grassroots democratic metaverse}}.
\newblock {\slshape \bibinfo{journal}{arXiv preprint arXiv:2203.04090}} \bibinfo{volume}{N/A}(\bibinfo{number}{N/A}), p. \bibinfo{pages}{N/A}, \doi{10.48550/arXiv.2203.04090}.

\bibitemdeclare{inproceedings}{shehory1998multi}
\bibitem{shehory1998multi}
\bibinfo{author}{Onn~M \surnamestart Shehory\surnameend}, \bibinfo{author}{Katia \surnamestart Sycara\surnameend} \& \bibinfo{author}{Somesh \surnamestart Jha\surnameend} (\bibinfo{year}{1998}): \emph{\bibinfo{title}{Multi-agent coordination through coalition formation}}.
\newblock In: {\slshape \bibinfo{booktitle}{Intelligent Agents IV Agent Theories, Architectures, and Languages: 4th International Workshop, ATAL'97 Providence, Rhode Island, USA, July 24--26, 1997 Proceedings 4}}, \bibinfo{publisher}{Springer}, \bibinfo{address}{N/A}, pp. \bibinfo{pages}{143--154}, \doi{10.1007/BFb0026756}.

\bibitemdeclare{article}{wang2016compare}
\bibitem{wang2016compare}
\bibinfo{author}{Shuohang \surnamestart Wang\surnameend} \& \bibinfo{author}{Jing \surnamestart Jiang\surnameend} (\bibinfo{year}{2016}): \emph{\bibinfo{title}{A compare-aggregate model for matching text sequences}}.
\newblock {\slshape \bibinfo{journal}{arXiv preprint arXiv:1611.01747}} \bibinfo{volume}{N/A}(\bibinfo{number}{N/A}), p. \bibinfo{pages}{N/A}, \doi{10.48550/arXiv.1611.01747}.

\bibitemdeclare{inproceedings}{wanyama2006negotiation}
\bibitem{wanyama2006negotiation}
\bibinfo{author}{Tom \surnamestart Wanyama\surnameend} \& \bibinfo{author}{Behrouz~H \surnamestart Far\surnameend} (\bibinfo{year}{2006}): \emph{\bibinfo{title}{Negotiation coalitions in group-choice multi-agent systems}}.
\newblock In: {\slshape \bibinfo{booktitle}{Proceedings of the Fifth International Joint Conference on Autonomous Agents and Multiagent Systems}}, \bibinfo{publisher}{N/A}, \bibinfo{address}{N/A}, pp. \bibinfo{pages}{408--410}, \doi{10.1145/1160633.1160704}.

\bibitemdeclare{article}{yang2024llm}
\bibitem{yang2024llm}
\bibinfo{author}{Joshua~C \surnamestart Yang\surnameend}, \bibinfo{author}{Marcin \surnamestart Korecki\surnameend}, \bibinfo{author}{Damian \surnamestart Dailisan\surnameend}, \bibinfo{author}{Carina~I \surnamestart Hausladen\surnameend} \& \bibinfo{author}{Dirk \surnamestart Helbing\surnameend} (\bibinfo{year}{2024}): \emph{\bibinfo{title}{Llm voting: Human choices and ai collective decision making}}.
\newblock {\slshape \bibinfo{journal}{arXiv preprint arXiv:2402.01766}}, \doi{10.1609/aies.v7i1.31758}.

\bibitemdeclare{article}{llm}
\bibitem{llm}
\bibinfo{author}{Wayne~Xin \surnamestart Zhao\surnameend}, \bibinfo{author}{Kun \surnamestart Zhou\surnameend}, \bibinfo{author}{Junyi \surnamestart Li\surnameend}, \bibinfo{author}{Tianyi \surnamestart Tang\surnameend}, \bibinfo{author}{Xiaolei \surnamestart Wang\surnameend}, \bibinfo{author}{Yupeng \surnamestart Hou\surnameend}, \bibinfo{author}{Yingqian \surnamestart Min\surnameend}, \bibinfo{author}{Beichen \surnamestart Zhang\surnameend}, \bibinfo{author}{Junjie \surnamestart Zhang\surnameend}, \bibinfo{author}{Zican \surnamestart Dong\surnameend} et~al. (\bibinfo{year}{2023}): \emph{\bibinfo{title}{A survey of large language models}}.
\newblock {\slshape \bibinfo{journal}{arXiv preprint arXiv:2303.18223}}, \doi{10.48550/arXiv.2303.18223}.

\bibitemdeclare{inproceedings}{zvi2021iterative}
\bibitem{zvi2021iterative}
\bibinfo{author}{Gil~Ben \surnamestart Zvi\surnameend}, \bibinfo{author}{Eyal \surnamestart Leizerovich\surnameend} \& \bibinfo{author}{Nimrod \surnamestart Talmon\surnameend} (\bibinfo{year}{2021}): \emph{\bibinfo{title}{Iterative Deliberation via Metric Aggregation}}.
\newblock In: {\slshape \bibinfo{booktitle}{International Conference on Algorithmic Decision Theory}}, \bibinfo{publisher}{Springer}, \bibinfo{address}{N/A}, pp. \bibinfo{pages}{162--176}, \doi{10.1007/978-3-030-87756-9_11}.

\end{thebibliography}

\appendix
\section{Missing Text}
\subsection{A Concrete Example}

Consider the following, toy example.

\begin{example}
Consider a metric space $X$ with a set of elements $P$ and a given distance $d$. We have a status quo $r\in P$ and three agents $A$, $B$, and $C$, each with its ideal point, $p^A$, $p^B$, $p^C$. Furthermore:
\begin{itemize}
    \item each agent is non-altruistic ($\sigma=0$);
    \item there is no coalition discipline;
    \item the mediator's $\alpha$ is set to be $0$;
    \item $\{p^A,p^B,p^C\}$ serve as the initial singleton coalition points.
\end{itemize}

The distances between the different ideal points of the agents and the status quo within the metric space are as follows (note that it is indeed a metric): $\mathbf{d}(p^A, p^A) = \mathbf{d}(p^B, p^B) = \mathbf{d}(p^C, p^C) = \mathbf{d}(p^D, p^D) = 0; \mathbf{d}(p^A, p^B) = 3; \mathbf{d}(p^A, p^C) = 5; \mathbf{d}(p^A, r) = 9; \mathbf{d}(p^B, p^C) = 2; \mathbf{d}(p^C, r) = 6; \mathbf{d}(p^C, r) = 8$. Consider another element of the metric space, $d^{BC}$, with $\mathbf{d}(p^B, p^{BC}) = \mathbf{d}(p^C, p^{BC}) = 1$.

% \[
% \begin{array}{ccccc}
% & A & B & C & r\\
% A & 0 & 3 & 5 & 9 \\
% B & & 0 & 2 & 6\\
% C & & & 0 & 8 \\
% r & & & & 0\\
% \end{array}
% \]

\begin{enumerate}
\item \textbf{Initialization:} Each agent starts with its own coalition.
\[
\begin{aligned}
&D = \{(C_A, p^A), (C_B, p^B), (C_C, p^C)\} \\
&C_A = \{A\}, \quad C_B = \{B\}, \quad C_C = \{C\}\ .
\end{aligned}
\]

\item \textbf{Iteration 1:} The mediator suggests the compromise point $p^{BC}$ to the coalitions $(C_B, p^B)$ and $(C_C, p^C)$. Both agents approve since $1 < 6$ and $1 < 8$. We arrive to the following coalition structure $D'$:
\[
\begin{aligned}
&D' = \{(C_A, p^A),(C_{BC}, p^{CB})\}\ , \\
&C_{BC} = \{B, C\}\ .
\end{aligned}
\]

\item \textbf{Halting condition:} A coalition with an agent majority has been formed (as $\nicefrac{|C_{BC}|}{|D|}>0.5$), thus the process halts.

\end{enumerate}
\end{example}

\subsection{Mediators in a 2D Euclidean Space}\label{section:euclidean}

In this section, we consider a rather simple setting where the metric space $X$ contains points in a 2D Euclidean space and the distance is $\ell_2$. This serves to illustrate the fundamental properties of our model and showcases the operation of our algorithms.
As a usecase, consider a scenario in which an agent community collaborates to mutually select a location for a social event (e.g., a picnic).

\subsubsection{Simulation-Based Analysis}

We describe the  design of the computer-based simulations we have conducted; and report and discuss the results.

We have generated instances of our model for the realization described above for a 2-dimensional Euclidean space. Next are details of the specific configuration used:
\begin{itemize}

\item Status Quo: Generated uniformly at random between $(0,200) \times (0,200)$.

\item Ideal Points: Drawing inspiration from the literature~\cite{elkind2017multiwinner}, each agent was assigned an ideal point $(x, y)$ with both coordinates sampled from either the uniform distribution between $(0,200)$ or from a 2-dimensional Gaussian Mixture Model (GMM). The GMM represents the overall probability distribution as a weighted sum of several Gaussian components with multiple peaks. In our simulations, we considered GMMs with $g$ combined Gaussian distributions, for $g \in \{0, 1, 2, 3, 4\}$ with the mean of each Gaussian being distributed uniformly between $0$ and $200$ in each dimension, its deviation distributed uniformly between $0$ and $50$, and the weights signifying the importance of each Gaussian are distributed from the $Dirichlet(\alpha^g \in {\mathbb{R}_0^+}^g)$ distribution with $\alpha$ set to $1$ (resulting in $g$ numbers that sum to $1$). This sampling of ideal points is demonstrated in the supplementary material. Note that we treat GMM with $g = 0$ (i.e., $0$ peaks) as the uniform distribution.

\item 
The different number $n$ of agents used in the simulations was $n \in \{10,20,30,40,50,100,250,1000\}$.

\item 
Coalition Discipline: we evaluated and compared instances with coalition discipline and without (as described in the realization Subsection in the main text.

\item For the mediator, we have used $\alpha=\{-1,0,1\}$ (as described in realization Subsection int he main text).

\item We have used $\sigma_v \in \{0,10,20,30\}$ as the degree of altruism, representing the smoothing of agents' approval functions (as described in the realization Subsection in the main text).

\item For the initialization of the singleton coalitions we set a parameter $I\in \{True, False\}$:
for $I = False$ the initial singleton coalition points were set to be the ideal points of each agent; while for $I=True$, the initial singleton coalition points were generated using a 2-dimensional Gaussian distribution with a mean being the ideal point $p^v$ and with a covariance matrix with $\sigma_x \sim U(0, 10)$ and $\sigma_y \sim U(0,10)$ on the diagonal (and zeros off-diagonal). 

\end{itemize}

We conducted $100$ independent repetitions for each configuration. Next we present our two evaluation metrics (the first measures the process speed, while the second measures the process quality):
\begin{itemize}

\item \textbf{Speed of convergence}: average number of iterations until the halting condition is met.
     
\item \textbf{Quality of converged state}: average distance between the proposal of the coalition containing an agent majority to the ideal points of the agents within that coalition; formally, for the single coalition $d = (C, p)$ in the halting state, with $\frac{|C|}{n}\geq 0.5$ we compute $\frac{1}{|C|}\sum_{v=1}^{|C|}\mathbf{d}(p,p^v)$.
     
\end{itemize}

For efficiency, we halt our simulations whenever the number of iterations exceeds a threshold of $10,000$ (i.e., we treat an instance for which no convergence is reached within $10,000$ iterations as an instance that does not converge at all).

\subsubsection{Results and Discussion}

Next we discuss the main conclusions, drawn at a $5\%$ significance level:
\begin{enumerate}
    \item Processes with coalition discipline and non-altruist agents agents always result in some non-convergences (i.e., the number of iterations is greater than 10,000) while all other combinations result in convergence.
    \item More agents result in more iterations (linearly), shorter mean distances, and higher log-odds of a converging process before 10,000 iterations.
    \item Higher $\alpha$ leads to a larger mean average distance.
    \item Coalition discipline shortens the mean average distance.
    \item High interaction between $n$ and $\alpha$, $n$ and coalition discipline, and $\sigma$ and coalition discipline results in more iterations until the halting condition.
    \item High interaction between $n$ and number of peaks (GMM), and coalition discipline and number of peaks (GMM), leads to fewer iterations until the halting condition.  
\end{enumerate}

\section{Illustrating the Simulation Ftramework}

\begin{example}
To better illustrate the process, we present one of the simulations conducted with fixed parameters outlined in Figure~\ref{Figure 3}. The simulation  involves 10 ideal sentences of agents regarding dealing with global warming (of maximum 15 words) projected onto a 2D Euclidean space, showcasing the coalitions each agent belongs to by the time the process concludes (i.e., the halting condition is satisfied). The visualization method employed for multi-dimensional data is adapted from~\cite{van2008visualizing}.

\begin{figure}[t]
  \centering
\includegraphics[width=0.7\linewidth]{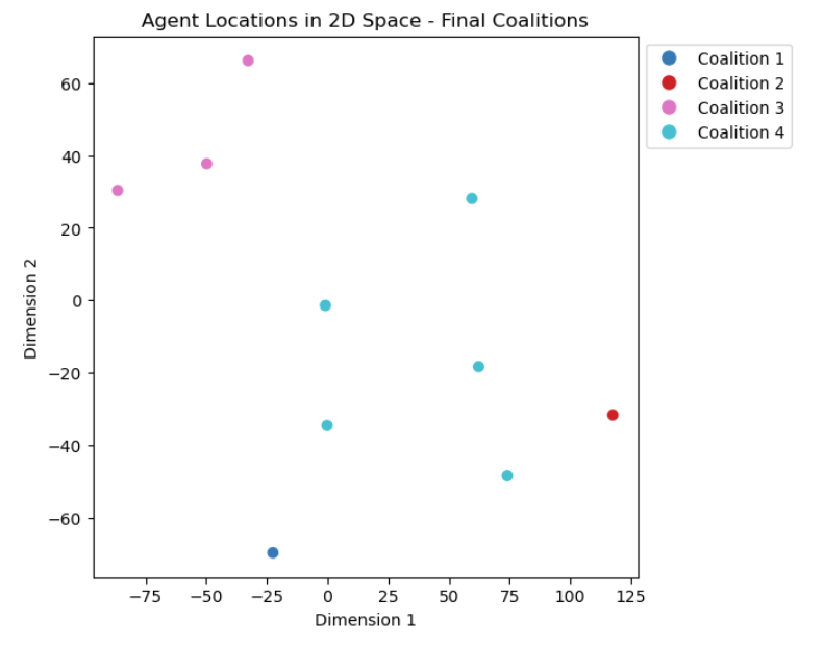}
  \caption{Coalition formation result- "Dealing with Global Warming", for  $n=10, C=\text{False}, \alpha=0, \sigma=0, I=\text{True}$.\vspace{15pt}}
  \label{Figure 3}
\end{figure}
\end{example}
Implementation, Code, and further Illustrations can be found here~\url{https://github.com/EyalBriman/AI-Mediator}.
\clearpage

\end{document}